\documentclass[aps,prd,preprint,groupedaddress]{revtex4}
\usepackage{amssymb}
\usepackage{amsmath}
\usepackage{graphicx} % Include figure files
\usepackage{dcolumn}  % Align table columns on decimal point
\usepackage{feynmf}   % Drawing Feynman Diagrams
\usepackage{epsfig}   % For PostScript figures
\newcommand{\Tr}{ {\mathbf{Tr}\, }}

\begin{document}

\preprint{}

%\title{Effective Locality: A New Property/Formulation of QCD?}
\title{Analytic, Non-Perturbative, Gauge-invariant QCD: Nucleon Scattering and Binding Potentials}


\author{H. M. Fried$^{\dag}$, Y. Gabellini$^{\ddag}$, T. Grandou$^{\ddag}$ and Y.-M. Sheu$^{\ddag}$}
\email[]{ymsheu@alumni.brown.edu}
%\thanks{}
\affiliation{${}^{\dag}$ {Physics Department, Brown University, Providence, RI 02912, USA} \\ ${}^{\ddag}$ {Universit\'{e} de Nice Sophia-Antipolis, Institut Non Lin$\acute{e}$aire de Nice, UMR 6618 CNRS, 06560 Valbonne, France}}

\date{\today}

\begin{abstract}
Removal of the quenched approximation in the mechanism which produced an analytic estimate of quark-binding potentials, along with a reasonable conjecture of the color structure of the nucleon formed by such a binding potential, is shown to generate an effective,  nucleon scattering and binding potential. The mass-scale factor on the order of the pion mass, previously introduced to define transverse imprecision of quark coordinates, is again used, while the strength of the potential is proportional to the square of a renormalized QCD coupling constant. The potential so derived does not include corrections due to spin, angular momentum, nucleon structure, and electroweak interactions; rather, it is qualitative in nature, showing how Nuclear Physics can arise from fundamental QCD.
\end{abstract}

%\keywords{}

\maketitle

\section{\label{SEC1}Introduction}

In previous papers~\cite{FootNote1,Fried2009_QCD1,Fried2010_QCD2,Fried2011_QCD3,Fried2011_QCD4} a new, analytic, non-perturbative, gauge-invariant approach to QCD has been defined and used to give a simple estimation of a quark-binding potential, $V(r)\simeq \xi \, \mu \, (r \, \mu)^{(1+\xi)}$, where $\xi$ is a small, real, positive parameter, $\xi \ll 1$, introduced phenomenologically into that function which guarantees transverse quark-coordinate imprecision; $\xi$ and $\mu$ are to be determined in terms of the pion and nucleon masses, as noted in Ref.~\cite{Fried2011_QCD4}; $\mu$ is a mass scale parameter understood to be on the order of the pion mass, $m_{\pi}$.

	In this paper, we begin with the concept of three bound quarks scattering against another triad of three bound quarks, with their full, non-perturbative exchanges of gluons between all the quarks taking place.  We assume that those triads that are initially bound remain bound at all times, which carries the implication that the multiple gluons exchanged between these nucleons do not change the overall color-singlet nature of each nucleon.  We neglect all electroweak interactions; and to further simplify the analysis assume that this nucleon scattering takes place at high relative velocities, so that a simplifying eikonal description of the scattering may be used.  We further simplify the analysis by neglecting spin effects -- which could be inserted if desired -- and aim for a simple, qualitative picture of how forces between nucleons can arise, starting from the basic fundamentals of QCD.

	In one sense, however, our eikonal model must be made more complicated than those quenched models used previously, for it turns out that one must here retain at least the simplest effects of the closed-quark-loop, or vacuum functional, $\mathbf{L}[A]$.  The Physics underlying this requirement follows because the forces which arise between quarks due to the multiple exchange of gluons are strong, tending to bind, for impact parameter $b$ separations on the order of $1/\mu$, but they fall off rapidly with increasing $b > 1/\mu$.  (At distances $b \ll 1/\mu$ large color fluctuations tend to reduce the value of any amplitude, and that could very well correspond to a non-perturbative translation of asymptotic freedom). How can nucleons, whose internal structures are defined at distances $b \sim 1/\mu$, then feel strong forces at separations $b > 1/\mu$?

	The answer is that vacuum loops, defined by $\mathbf{L}[A]$, can stretch in the transverse directions, and can serve to transmit the multiple gluon interactions across larger values of impact parameter; a gluon "bundle" from one nucleon attaches itself to one point on the loop, while another bundle of gluons passes from a second point on the loop (at a significant transverse distance from the first) to the other nucleon.  Although this passage of momentum via a closed vacuum loop changes the interaction somewhat, an essential "short-range" interaction is produced at distances larger than would be possible by quark-to-quark passage alone.  And if, as this loop is stretched in a transverse direction, one transverse side of the loop corresponds to a quark and the other to an antiquark, one has the image of an effective pion being exchanged between scattering nucleons.

	Effective Locality (EL) is the reason that multiple gluon exchange appears to leave or arrive at a single space-time point on a "quark line", in the "bundle-diagrams" to follow, which correspond to the sums over infinite numbers of conventional Feynman graphs.  Such sums are possible because they can be represented by the equivalent of Gaussian functional integration over products of factors of $\mathbf{G}_{\mathrm{c}}[A]$ and $\mathbf{L}[A]$, where $\mathbf{G}_{\mathrm{c}}[A]$ represents a quark propagating in a "classical" gluon vector potential $A_{\mu}^{a}(x)$; and that integration can be carried out exactly because there exist Fradkin representations of these two functionals which are Gaussian in their $A$-dependence.  Results are then expressed in terms of these Fradkin representations, which are just Potential Theory constructs, and have relatively simple approximations in different physical situations.  In this paper, for reasons of simplicity, we shall replace the $\mathbf{G}_{\mathrm{c}}[A]$ of each quark in a nucleon, and the single $\mathcal{G}_{\mathrm{c}}[A]$ which models that nucleon, by its high-energy eikonal limit, and then connect the gluon bundles emitted by each nucleon to two, and only two points on a single loop. More complicated loop structures are certainly possible, and should be investigated, but this is the simplest representation of "effective pion exchange between nucleons".

	Finally, mention should be made of the relative simplicity of this approach, compared to other well-known and long-studied methods of calculation in QCD.  Again, it is the unexpected EL, appearing automatically after the non-perturbative sum over all possible, gauge-invariant, gluon exchanges between quarks has occurred, which is responsible for the huge simplifications obtained as Halpern's functional integrals are here reduced to a finite set of ordinary integrals \cite{bctg} amenable to computer evaluation, and here estimated in the simplest way possible.  One has long believed in the Principle of "\emph{Conservation of Difficulty}", when calculating higher-order effects in QED or any Abelian theory; but for non-Abelian field theories, approached in the manner we suggest, that Principle is not true.  %There is always an exception to every long-held principle or belief; "\emph{there's a crack in everything, that's how the light gets in.}"~\cite{LeonardCohen}

\section{\label{SEC2}Formulation}

We begin by first presenting a bird's eye view of the detailed calculations that follow immediately, expressed in terms of the analysis of \cite{Fried2009_QCD1}, and in particular to the discussion  centering about its Eqs. (20) - (23), where the color coordinates of each of the two $\mathbf{G}_{\mathrm{c}}[A]$, each representing a scattering quark or antiquark, are discussed, in connection with the eikonal forms at high energies, taken by the exact Fradkin representation of that Green's function.  Now generalize that process to the scattering of a triad of quarks bound into a singlet nucleon with another such nucleon.  For each nucleon, there will occur the product of three such terms as written in Eq. (23) of \cite{Fried2009_QCD1}, each with the same fraction of that nucleon's momentum, and with their color weightings $\Omega^{a}$ restricted so as to insure that the three bound quarks comprising each nucleon remain in a color-singlet state.

	We replace the description of that combination by that of a nucleon, of momentum $p$ and effective color weighting $\bar{\Omega}^{a}$, so defined such that only those combinations of Gell-Mann matrices of each of the basic quarks corresponds to gluons which may be exchanged and so preserve each triad of quarks in its bound, color-singlet state.  Note that all such color-singlet exchanges can be absorbed and emitted by the quark line comprising the loop $\mathbf{L}[A]$, for the Fradkin representation of $\mathbf{L}[A]$ contains a trace over all possible combinations of color coordinates.  By this simplification, we replace the essence of a 6-body problem by a 2-body problem; and this has the consequence that our subsequent estimation of the nucleon-nucleon binding potential produces a qualitative description of how nuclear forces can arise from basic QCD.

%By this simplification, we replace the essence of a 6-body problem by a 2-body problem; and this is possible because, and has the consequence that our subsequent estimation of the qualitative scattering potential suppresses the precise color dependence of the result.  That dependence modifies the magnitude and possibly the shape of the resulting potential, and remains to be calculated properly; but such details are not the aim of this paper, which is to produce a qualitative description of how nuclear forces can arise from basic QCD.

	We next refer the reader to Section IV of Ref.~\cite{Fried2010_QCD2}, and in particular to the functional operations of its Eq.~(30), with attention drawn to the translation operator $\exp[-\int{\mathcal{Q} \cdot (g f \cdot \chi)^{-1} \cdot \frac{\delta}{\delta A}}]$ acting upon $\exp{\{\mathbf{L}[A]\}}$.  Here, $\mathcal{Q}$ refers to the coordinates of the two nucleons, and the translation operator inserts that dependence in a well-defined way into the $\mathbf{L}[A]$ written in Eq.~(A11) of that paper.  The remaining linkage operator of (30) generates a functional cluster expansion, discussed and derived in Ref.~\cite{HMF1} and \cite{HMF2}; for our purposes, involving but a single loop and suppressing any gluon bundles exchanged across that loop, this remaining linkage operation can be neglected.  This simplified analysis, which will require a simple renormalization, is sufficient to  produce a reasonable, qualitative nuclear potential from basic 'realistic' QCD.  By `realistic', it meant a formulation of QCD which contains from its inception that asymptotic quarks and/or antiquarks are found only in bound states, and hence their transverse coordinates cannot be specified exactly.  In this paper, such transverse imprecision follows from the defining arguments found in \cite{Fried2011_QCD3}.

	The specific steps of this analysis follow from a statement of the Generating Functional (GF) derived in \cite{Fried2009_QCD1},
\begin{eqnarray}\label{Eq:1}
\mathfrak{Z}_{\mathrm{QCD}}^{(\zeta)}[j,\bar{\eta}, \eta] &=& \mathcal{N} \, \int{\mathrm{d}[\chi] \, e^{ \frac{i}{4} \int{ (\chi_{\mu \nu}^{a})^{2} }} \, e^{\frac{i}{2} \, \int{j_{\mu}^{a} \, \mathbf{D}_{\mathrm{c},\mu\nu}^{ab} \, j_{\nu}^{b}}} \,} \\ \nonumber & & \left. e^{\mathfrak{D}_{A}} \cdot e^{\frac{i}{2} \int{\chi_{\mu\nu}^{a} \, \mathbf{F}_{\mu \nu}^{a} + \frac{i}{2} \int{ A_{\mu}^{a} \, \left.\left( \mathbf{D}_{\mathrm{c}}\right)^{-1}\right|_{\mu \nu}^{a b} \, A}_{\nu}^{b} }} \cdot e^{i\int{\bar{\eta} \cdot \mathbf{G}_{\mathrm{c}}[A] \cdot \eta} + \mathbf{L}[A]}\right|_{A = \int{\mathbf{D}_{\mathrm{c}} \cdot j} },
\end{eqnarray}

\noindent which is nothing else but a particular, and gauge-invariant rearrangement of the Schwinger/Symanzik GF derived more than a half-century ago, and applied to QCD.  (Such a representation of gauge invariance is dependent upon cubic and quartic gluon interactions, and is not possible for QED.)  Here, the covariant, causal gluon propagator is given, in momentum space, by
\begin{equation}
\tilde{\mathbf{D}}_{\mathrm{c},\mu\nu}^{ab} = \frac{\delta^{ab}}{k^{2} - i\epsilon} \, \left[ g_{\mu \nu} - \zeta \frac{k_{\mu} k_{\nu}}{k^{2} - i\epsilon}\right], \quad k^{2} = \vec{k}^{2} - k_{0}^{2},
\end{equation}

\noindent with $\zeta$ an arbitrary gauge parameter, while $\exp{\{\mathfrak{D}_{A}\}}$ represents the "linkage operator", with
\begin{equation}
\mathfrak{D}_{A} =  - \frac{i}{2} \int{\frac{\delta}{\delta A_{\mu}^{a}} \cdot  \mathbf{D}_{\mathrm{c},\mu\nu}^{ab} \cdot \frac{\delta}{\delta A_{\nu}^{b}} },
\end{equation}

\noindent This functional differentiation formalism has the great advantage over the often more customary Functional Integral (FI) over gluon fluctuations, in that there is no need for concern about spurious (Gribov) replication of gauge copies; it is gauge-invariant\cite{FootNote3} from the very beginning, and made so by means of a small observation overlooked for decades, as described in~\cite{Fried2009_QCD1}.

	The functional $\mathbf{G}_{\mathrm{c}}[A] = \left[ m + \gamma_{\mu} \, (\partial_{\mu} - i g A_{\mu}^{a} \lambda^{a})\right]^{-1}$ represents a quark propagating in the presence of a specified, "classical" field $A_{\mu}^{a}(x)$, while $\mathbf{L}[A] = \Tr{\ln{\left[ 1 - ig (\gamma \cdot A \cdot \lambda) \, \mathbf{S}_{\mathrm{c}} \right]}}$, $\mathbf{S}_{\mathrm{c}}  = \mathbf{G}_{\mathrm{c}}[0]$, denotes the "vacuum" functional; both functionals have Fradkin representations essentially Gaussian in their $A$-dependence, which permits the linkage operations needed for the specific amplitudes desired (obtained by functional differentiation with respect to the sources $\eta(x)$, $\bar{\eta}(y)$, $j_{\mu}^{a}(z)$) to be obtained exactly; and this, in turn, corresponds to the summation of an infinite number of classes of Feynman graphs, each class containing an infinite number of graphs, with the results expressed in terms of the parameters of the Fradkin representations for $\mathbf{G}_{\mathrm{c}}$ and $\mathbf{L}$.  In this paper, for simplicity, we shall replace the Fradkin representation for $\mathbf{G}_{\mathrm{c}}[A]$ by its high-energy, eikonal approximation; and, again for simplicity, restrict consideration of $\mathbf{L}[A]$ to its simplest form in which quark spins have been suppressed. Finally, as discussed and derived in~\cite{Fried2011_QCD3}, the "transverse averaging" needed for "realistic" QCD, after the non-perturbative sums over Feynman graphs have been computed, can be very simply inserted as an intermediate step by the replacement
\begin{equation}
A_{\mu}^{a}(x-u(s')) \rightarrow \int{\mathrm{d}^{2}x'_{\perp} \, \mathfrak{a}(\vec{x}_{\perp} - \vec{x}'_{\perp}) \, A_{\mu}^{a}(x'-u(s')) }, \quad x'_{\mu} \equiv (ix_{0}; \vec{x}'_{\perp},x_{\mathrm{L}}),
\end{equation}

\noindent where $\mathfrak{a}(x_{\perp}  - x'_{\perp})$ is a real, symmetric function, expressing the impossibility of defining precise transverse coordinates of bound quarks and/or antiquarks.  (Originally, as in~\cite{Fried2011_QCD3}, only the quark color current operator was endowed with such transverse imprecision; but because of the assumed symmetry of the $\mathfrak{a}(x_{\perp}  - x'_{\perp})$, in that part of the Action operator coupling such color current to gluons, it is permissible to interchange the roles of transversally-shifted fields, and imagine that it is the coordinate dependence in $A$ which has been shifted.  In reality, no such shift of the $\mathbf{F}_{\mu \nu}^{a}(x)$ have been performed.)

	It may also be noted that cubic and quartic gluon interactions are fully included in this formalism, and are represented by the Halpern FI over $\mathrm{d}[\chi]$.  And because of the remarkable property of EL, alluded to above, which appears after the non-perturbative sums are performed, the Halpern FI of (\ref{Eq:1}) reduces to a finite set of ordinary integrals \cite{bctg}, which are amenable to numerical computation.  In this paper, again for simplicity, we estimate these integrals by means of an approximate Gaussian evaluation.

	We now ask the reader to imagine that functional derivatives are taken with respect to six pairs of $\eta$, $\bar{\eta}$ sources, "bringing down" six $\mathbf{G}_{\mathrm{c}}(x^{(i)}, y^{(i)}|A)$, which we divide into two groups of three,
\begin{equation}
\mathbf{G}_{\mathrm{c}}(x_{\mathrm{I}}^{(1)}, y_{\mathrm{I}}^{(1)}|A) \,\mathbf{G}_{\mathrm{c}}(x_{\mathrm{I}}^{(2)}, y_{\mathrm{I}}^{(2)}|A) \, \mathbf{G}_{\mathrm{c}}(x_{\mathrm{I}}^{(3)}, y_{\mathrm{I}}^{(3)}|A)
\end{equation}

\noindent for nucleon $\mathrm{I}$, and a similar triad with coordinate superscripts (4), (5), and (6) for nucleon $\mathrm{I\!I}$, beginning the computation as if we were calculating a six-quark scattering amplitude.  Each $\mathbf{G}_{\mathrm{c}}[A]$ will bring to its triad the $A$-dependence associated with an eikonal/high-energy limit of its exact Fradkin representation, of form
\begin{equation}\label{Eq:2}
\exp{\left[ -ig \frac{p_{\mu,(\mathrm{I},\mathrm{I\!I})}^{(i)}}{m} \, \int_{-\infty}^{+\infty}{\mathrm{d}s \, \Omega_{(\mathrm{I},\mathrm{I\!I})}^{a, (i)}(s) \, A_{\mu}^{a}(y_{(\mathrm{I},\mathrm{I\!I})}^{(i)} - s \, \frac{p_{(\mathrm{I},\mathrm{I\!I})}^{(i)}}{m}) }\right]}.
\end{equation}

\noindent We now introduce the bound-state nature of each triad of quarks by first suppressing the coordinate superscripts, $y_{(\mathrm{I},\mathrm{I\!I})}^{(i)} \rightarrow y_{(\mathrm{I},\mathrm{I\!I})}^{}$, and $p_{(\mathrm{I},\mathrm{I\!I})}^{(i)} \rightarrow (1/3) \, p_{(\mathrm{I},\mathrm{I\!I})}^{}$, since each quark must have essentially the same space-time and momentum coordinates if its nucleon is to remain intact.  This means that the product of the three factors of Eq.~(\ref{Eq:2}) which are now a property of each nucleon may be written as
\begin{equation}\label{Eq:3}
\exp{\left[ -i\frac{g}{3} \frac{p_{\mu,(\mathrm{I},\mathrm{I\!I})}^{}}{m} \, \int_{-\infty}^{+\infty}{\mathrm{d}s \, \sum_{i=1}^{3}{\Omega_{(\mathrm{I},\mathrm{I\!I})}^{a, (i)}(s) \, A_{\mu}^{a}(y_{(\mathrm{I},\mathrm{I\!I})}^{} - \frac{s}{3} \, \frac{p_{(\mathrm{I},\mathrm{I\!I})}^{}}{m}) }}\right]}.
\end{equation}

	We emphasize that (\ref{Eq:3}) refers to the $A$-dependence of each nucleon after the linkage operations binding each triad of quarks have been performed, as discussed in~\cite{Fried2011_QCD4}, which analysis we here suppress.  Further, for each nucleon to remain bound for all (proper) times, there must exist a relation between the $\Omega^{a, (i)}(s)$ such that only color singlets are exchanged between nucleons $\mathrm{I}$ and $\mathrm{I\!I}$, and this relation should be independent of $s$.  Since the $\Omega^{a, (i)}$ define the Gell-Mann matrices $\lambda^{a, (i)}$ inserted between initial and final nucleon states, there must be a relation between the $\Omega^{a, (i)}$ guaranteeing that each nucleon remains a color singlet.  We thus simplify Eq.~(\ref{Eq:3}) by introducing $\bar{\Omega}_{(\mathrm{I},\mathrm{I\!I})}^{a} = \sum_{i=1}^{3}{\Omega_{(\mathrm{I},\mathrm{I\!I})}^{a, (i)}}$, and re-scaling $s \rightarrow 3s$, so that Eq.~(\ref{Eq:3}) becomes
\begin{equation}\label{Eq:4}
\exp{\left[ -ig \frac{p_{\mu,(\mathrm{I},\mathrm{I\!I})}^{}}{m} \, \bar{\Omega}_{(\mathrm{I},\mathrm{I\!I})}^{a} \, \int_{-\infty}^{+\infty}{\mathrm{d}s \, A_{\mu}^{a}(y_{(\mathrm{I},\mathrm{I\!I})}^{} - s \, \frac{p_{(\mathrm{I},\mathrm{I\!I})}^{}}{m}) }\right]}.
\end{equation}

	A modification, representing the "realistic" QCD defined and used in the two preceding papers~\cite{Fried2011_QCD3,Fried2011_QCD4}, replaces in the exact Fradkin representation each $A_{\mu}^{a}(y'-u(s'))$ by
\begin{equation}\label{Eq:5}
\int{\mathrm{d}^{2}y'_{\perp} \, \mathfrak{a}(y_{\perp}^{} - y'_{\perp}) \, A_{\mu}^{a}(y'-u(s'))},
\end{equation}

\noindent where $y'_{\mu} = (i y_{0}^{}; \vec{y}_{\perp}^{\, '}, y_{\mathrm{L}}^{})$ and $y_{\mu}^{} = (i y_{0}^{}; \vec{y}_{\perp},y_{\mathrm{L}}^{})$ represents the coordinate of an individual nucleon.  Since we are assuming strict binding of each nucleon, the only transverse imprecision we must specify for this analysis is that between the quarks of one nucleon and those of the other; and since we have assumed that such gluon exchanges are not in any way to disrupt the binding of quarks within each nucleon, we shall invoke transverse imprecision for the $y'_{\mu}$ coordinates of $A_{\mu}^{a}(y' - u(s'))$ of (\ref{Eq:5}), replacing (\ref{Eq:4}) by
\begin{equation}\label{Eq:6}
\exp{\left[ -i \int{\mathrm{d}^{4}w \, \mathcal{Q}_{\mu, (\mathrm{I}, \mathrm{I\!I})}^{a}(w) \,  A_{\mu}^{a}(w)} \right]}
\end{equation}

\noindent where
\begin{eqnarray}
\mathcal{Q}_{\mu, (\mathrm{I}, \mathrm{I\!I})}^{a}(w) = & & g \, \left( \frac{p_{\mu,(\mathrm{I},\mathrm{I\!I})}^{}}{m} \right) \, \bar{\Omega}_{(\mathrm{I},\mathrm{I\!I})}^{a} \, \\ \nonumber & & \quad \times \int{\mathrm{d}^{2}y'_{\perp} \, \mathfrak{a}(y_{\perp,(\mathrm{I},\mathrm{I\!I})}^{} - y'_{\perp,(\mathrm{I},\mathrm{I\!I})}) \, \delta^{(4)}(w - y'_{(\mathrm{I},\mathrm{I\!I})} + s \, \frac{p_{(\mathrm{I},\mathrm{I\!I})}^{}}{m})}.
\end{eqnarray}

\noindent In so doing, we have replaced the Fradkin coordinates $u_{\mu}(s)$ by the eikonal combinations $\frac{s \, p_{\mathrm{I}}^{}}{m}$ or $\frac{s \, p_{\mathrm{I\!I}}^{}}{m}$, and have neglected quark spin dependence.

	The binding process effectively transforms each triad of quark Green's functions into a single nucleon Green's function of mass $M$ and 4-momentum $p_{\mu}$ where the exponential of (\ref{Eq:6}), the effective `relic' of its original three quarks, is retained for subsequent use in calculating the interaction between both nucleons.  Each nucleon's Green's function now contributes to the eikonal scattering amplitude of the two nucleons, and in their CM takes on the standard form~\cite{Eikonal}
\begin{eqnarray}
\mathcal{T}(s,t) = \frac{is}{2M^{2}} \int{\mathrm{d}^{2} \, e^{i \vec{q} \cdot \vec{b}} \, \left[ 1 - e^{i \mathbb{X}(b,s)} \right] },
\end{eqnarray}

\noindent where $\vec{q}$ is the momentum transfer of this scattering process, $s = - (p_{\mathrm{I}}^{} + p_{\mathrm{I\!I}}^{})^{2}$, $t = - (p_{\mathrm{I}}^{} - p'_{\mathrm{I\!I}})^{2} = - \vec{q}^{\, 2}$, and where
\begin{eqnarray}\label{Eq:7}
e^{i \mathbb{X}(b,s)} = & & \mathcal{N} \, \int{\mathrm{d}[\chi] \, e^{\frac{i}{4} \int{\chi^{2}}} \, }\\ \nonumber & & \quad \cdot \left. e^{\mathfrak{D}_{A}} \cdot e^{\frac{i}{4} \int{\chi \cdot \mathbf{F}} + \frac{i}{2}\int{A \cdot (\mathbf{D}_{\mathrm{c}}^{-1}) \cdot A}} \, e^{-i \int{(\mathcal{Q}_{\mathrm{I}} + \mathcal{Q}_{\mathrm{I\!I}} )\cdot A}} \, e^{\mathbf{L}[A]} \right|_{A \rightarrow 0}
\end{eqnarray}

\noindent with normalization constant $\mathcal{N}$ defined such that $\mathbb{X} \rightarrow 0$ for $g \rightarrow 0$.

	The linkage operation of (\ref{Eq:7}) then has the Gaussian form
\begin{eqnarray}\label{Eq:8}
\left. e^{-\frac{i}{2}\int{\frac{\delta}{\delta A} \cdot \mathbf{D}_{\mathrm{c}} \cdot \frac{\delta}{\delta A}}} \cdot e^{ \frac{i}{2}\int{A \cdot \mathcal{K} \cdot A} + i \int{\mathcal{R} \cdot A}} \, e^{\mathbf{L}[A]} \right|_{A \rightarrow 0},
\end{eqnarray}

\noindent where $\mathcal{K}_{\mu\nu}^{ab} = g f^{abc} \chi_{\mu\nu}^{c} + (\mathbf{D}_{\mathrm{c}}^{-1})_{\mu\nu}^{ab}$ and $\mathcal{R}_{\mu}^{a} = \partial_{\nu} \chi_{\mu\nu}^{a} - \mathcal{Q}_{\mathrm{I},\mu}^{a} - \mathcal{Q}_{\mathrm{I\!I},\mu}^{a}$.  As in the passage from (21) to (31) of Ref.~\cite{Fried2010_QCD2}, the functional operation may be carried through exactly, yielding for (\ref{Eq:8}),
\begin{eqnarray}\label{Eq:9}
& & e^{-\frac{i}{2}\int{\mathcal{R} \cdot (g f \cdot \chi)^{-1} \cdot \mathcal{R} } + \frac{1}{2} \Tr{\ln{(g f \cdot \chi)^{-1}}}} \\ \nonumber & & \cdot \left. e^{+\frac{i}{2}\int{\frac{\delta}{\delta A} \cdot (g f \cdot \chi)^{-1} \cdot \frac{\delta}{\delta A}}} \cdot e^{- \int{\mathcal{R} \cdot (g f \cdot \chi)^{-1} \cdot \frac{\delta}{\delta A}}} \cdot e^{\mathbf{L}[A]} \right|_{A \rightarrow 0}.
\end{eqnarray}

	The first line of (\ref{Eq:9}) may be rewritten as
\begin{eqnarray}\label{Eq:10}
\left[ \det{(g f \cdot \chi)} \right]^{-\frac{1}{2}} \cdot \exp{\left\{ -\frac{i}{2} \, \int{\left[\partial \chi - \mathcal{Q}_{\mathrm{I}} - \mathcal{Q}_{\mathrm{I\!I}} \right] \cdot (g f \cdot \chi)^{-1} \cdot \left[\partial \chi - \mathcal{Q}_{\mathrm{I}} - \mathcal{Q}_{\mathrm{I\!I}} \right] } \right\}},
\end{eqnarray}

\noindent and we here rely on the strong coupling limit of $g \gg 1$, keeping only the terms $\mathcal{Q}_{\mathrm{I}, \mathrm{I\!I}}$ (this is really not necessary, but it simplifies the analysis; if the $\partial\chi$-terms are retained, the normalization integrals become more complicated, but the thrust of the procedure is the same).

	Furthermore, the terms of (\ref{Eq:10}) proportional to two factors of $\mathcal{Q}_{\mathrm{I}}$ and to two factors of $\mathcal{Q}_{\mathrm{I\!I}}$ are "self-energy" corrections to the respective nucleon propagators, and they will be suppressed, since we are here interested only in the interaction of one nucleon upon the other.  A similar remark may be made for those terms containing a single factor of $\partial \chi$ and either $\mathcal{Q}_{\mathrm{I}}$ or $\mathcal{Q}_{\mathrm{I\!I}}$, for they correspond to "tadpole"-like structures attached to either nucleon, and are not relevant here.  With these simplifications, (\ref{Eq:10}) is replaced by
\begin{eqnarray}\label{Eq:11}
\left[ \det{(g f \cdot \chi)} \right]^{-\frac{1}{2}} \cdot \exp{\left\{ -i \, \int{ \mathcal{Q}_{\mathrm{I}} \cdot (g f \cdot \chi)^{-1} \cdot \mathcal{Q}_{\mathrm{I\!I}}  } \right\}},
\end{eqnarray}

\noindent which, except for different color factors, has the form of the eikonal function describing the interaction between a pair of quarks.

\par
For the impact parameter range between nucleons in which we are interested, it turns out that (\ref{Eq:11}) gives an unimportant contribution to the nucleon-nucleon potential; for simplicity, we here neglect it, in contrast to the true source of that potential, which arises from the action of the linkage/displacement operators of (\ref{Eq:9}) upon $\exp{\{\mathbf{L}[A]\}}$. It is worth mentioning that this makes for an important difference with the quark binding potential evaluated in Ref.~\cite{Fried2011_QCD4}.

	Denoting the linkage operator of (\ref{Eq:9}) by
\begin{equation}
\exp{\left[\bar{\mathfrak{D}}_{A} \right]}  = \exp{\left[ -\frac{i}{2}\int{\frac{\delta}{\delta A} \cdot (-g f \cdot \chi)^{-1} \cdot \frac{\delta}{\delta A}}\right]},
\end{equation}

\noindent where $(-g f \cdot \chi)^{-1}$ represents each non-perturbative gluon bundle, as described in Ref.~\cite{Fried2011_QCD3}, to be exchanged between the quark lines which form the closed loop $\mathbf{L}[A]$, its action upon $\mathbf{L}[A]$ is most conveniently described in terms of a functional cluster decomposition as
\begin{eqnarray}\label{Eq:12}
e^{\bar{\mathfrak{D}}_{A}} \cdot e^{\mathbf{L}[A]} = \exp{\left[ \sum_{n=1}^{\infty}{\frac{1}{n!} \, \bar{\mathbf{L}}_{n}}  \right]}, \quad \bar{\mathbf{L}}_{n} = \left. e^{\bar{\mathfrak{D}}_{A}} \cdot (\mathbf{L}[A])^{n} \right|_{\mathrm{connected}},
\end{eqnarray}

\noindent where "connected" requires at least one gluon bundle exchanged between different $\mathbf{L}[A]$'s~\cite{ClusterExp}.  In this paper we shall be concerned only with the simplest possible application of a single closed loop, and for this we may suppress the linkage operation of (\ref{Eq:12}), while retaining the functional displacement operation of (\ref{Eq:9}).  With these simplifications, the second line of (\ref{Eq:9}) becomes
\begin{eqnarray}
\left. \exp{\left\{\mathbf{L}[A -  (g f \cdot \chi)^{-1} \cdot (\mathcal{Q}_{\mathrm{I}} + \mathcal{Q}_{\mathrm{I\!I}})] \right\}} \right|_{A \rightarrow 0} = \exp{\left\{\mathbf{L}[- (g f \cdot \chi)^{-1} \cdot (\mathcal{Q}_{\mathrm{I}} + \mathcal{Q}_{\mathrm{I\!I}})] \right\}},
\end{eqnarray}

\noindent and our eikonal simplifies to
\begin{eqnarray}\label{Eq:13}
e^{i \mathbb{X}(b,s)} = \mathcal{N} \, \int{\mathrm{d}[\chi] \, e^{\frac{i}{4} \int{\chi^{2}}} \cdot [\det{(g f \cdot \chi)}]^{-\frac{1}{2}} \cdot e^{\mathbf{L}[- (g f \cdot \chi)^{-1} \cdot (\mathcal{Q}_{\mathrm{I}} + \mathcal{Q}_{\mathrm{I\!I}})]} }.
\end{eqnarray}

	In order to calculate the vacuum loop contribution to (\ref{Eq:13}), we first write a Fradkin Representation for $\mathbf{L}[A]$, as in Ref.~\cite{Fried2010_QCD2},
\begin{eqnarray}\label{Eq:14}
\mathbf{L}[A] &=&  - \frac{1}{2} \Tr{} \int_{0}^{\infty}{\frac{dt}{t} \, e^{-it m^{2}}} \, \mathcal{N}(t) \, \int{d[v] \, e^{ \frac{i}{2} \int_{0}^{t}{dt' \, v \cdot (2h)^{-1} \cdot v } } \, \delta^{(4)}(v(t)) } \\ \nonumber && \quad \times \, \int{d^{4}x \, \int{\mathrm{d}[\hat{\alpha}] \, \int{\mathrm{d}[\hat{\Omega}] }}} \, e^{i \int{dt' \, \hat{\Omega}^{a}(t') \, \hat{\alpha}^{a}(t')} }  \, \left( e^{ i \int_{0}^{t}{dt' \, \hat{\alpha}^{b}(t') \, \lambda^{b}}} \right)_{+} \\ \nonumber & & \quad \times  \left[ e^{- i g \int_{0}^{t}{dt' \, v'_{\alpha}(t') \, \hat{\Omega}^{a}(t') \, \int{\mathrm{d}^{2}x_{\perp} \, \mathfrak{a}(x_{\perp} - x'_{\perp}) \, A^{a}_{\alpha}(x' - v(t'))} }} - 1 \right],
\end{eqnarray}

\noindent where $x'_{\mu} = (i x_{0}^{}; \vec{x}_{\perp}^{\, '}, x_{\mathrm{L}}^{})$, $h(t_{1}, t_{2}) = \theta(t_{1} - t_{2}) t_{2} + \theta(t_{2} -  t_{1})  t_{1} = \frac{1}{2}(t_{1} + t_{2} - |t_{1} - t_{2}|)$, $\mathcal{N}(t)$ is the normalization for the Gaussian functional integral over $v_{\alpha}(t')$, $\Tr{}$ denotes a trace over Dirac and color indices, and the hat notation of $\hat{\alpha}$ and $\hat{\Omega}$ is used to distinguish these loop color-variables from those of nucleons $\mathrm{I}$ and $\mathrm{I\!I}$.  Again, in the interests of simplicity, we shall neglect all spin dependence of the quark loop, and, for clarity, have chosen the longitudinal and transverse directions of the loop to lie in the respective directions defined by the nucleons in their CM.

	With the simplifications of the last two paragraphs, all of the structure that remains in our eikonal amplitude arises from that nucleon dependence, $\mathcal{Q}_\mathrm{I}$ and $\mathcal{Q}_\mathrm{I\!I}$, which has been translated into the argument of $\mathbf{L}$ in (\ref{Eq:13}), as its argument $A$ is shifted to  $-(g f \cdot \chi)^{-1} \cdot (\mathcal{Q}_\mathrm{I} + \mathcal{Q}_\mathrm{I\!I})$.  But this shift occurs in the exponential factor of (\ref{Eq:14}), whose expansion corresponds to multiple quark loops exchanged between the nucleons.  The simplest, and probably the most important effect arises from the exchange of a single quark loop, proportional to the factor $\mathcal{Q}_\mathrm{I}$ multiplying $\mathcal{Q}_\mathrm{I\!I}$, which may be extracted from the quadratic expansion of that exponential factor, neglecting tadpole and self-energy corrections to the nucleons.  We therefore replace the third line of (\ref{Eq:14}) by
\begin{eqnarray}\label{Eq:15}
& & + g^{2} \, \int_{-\infty}^{+\infty}{\mathrm{d}s_{1} \, \left( \frac{p_{\mathrm{I},\mu}^{}}{m}\right) } \,  \int_{-\infty}^{+\infty}{\mathrm{d}s_{2} \, \left( \frac{p_{\mathrm{I\!I},\nu}^{}}{m}\right) } \, \bar{\Omega}^{a} \, \bar{\Omega}^{b} \, \int_{0}^{t}{\mathrm{d}t_{1} \, v'_{\alpha}(t_{1}) \hat{\Omega}^{c}(t_{1})} \, \int_{0}^{t}{\mathrm{d}t_{2} \, v'_{\beta}(t_{2}) \hat{\Omega}^{d}(t_{2})} \\ \nonumber & & \quad \cdot \int{\mathrm{d}^{2}x'_{\perp} \, \mathfrak{a}(x_{\perp} - x'_{\perp}) \, \int{\mathrm{d}^{2}x''_{\perp} \, \mathfrak{a}(x_{\perp} - x''_{\perp}) }} \, \int{\mathrm{d}^{2}y'_{\mathrm{I},\perp} \, \mathfrak{a}(y_{\mathrm{I},\perp} - y'_{\mathrm{I},\perp})} \, \int{\mathrm{d}^{2}y'_{\mathrm{I\!I},\perp} \, \mathfrak{a}(y_{\mathrm{I\!I},\perp} - y'_{\mathrm{I\!I},\perp})} \\ \nonumber & & \quad \quad \cdot \delta^{(4)}(y'_{\mathrm{I}} - s_{1} \frac{p_{\mathrm{I}}}{m} - x' +v(t_{1})) \cdot \delta^{(4)}(y'_{\mathrm{I\!I}} - s_{2} \frac{p_{\mathrm{I\!I}}}{m} - x'' +v(t_{2})) \\ \nonumber & & \quad \quad \quad  \cdot \left. \left( f \cdot \chi(y'_{\mathrm{I}} - s_{1} \frac{p_{\mathrm{I}}}{m}) \right)^{-1} \right|_{\mu \alpha}^{a c} \cdot \left. \left( f \cdot \chi(y'_{\mathrm{I\!I}} - s_{2} \frac{p_{\mathrm{I\!I}}}{m}) \right)^{-1} \right|_{\beta \nu}^{d b},
\end{eqnarray}

\noindent where the generic notation of $z'_{\mu} = (iz_{0}; z'_{\perp},z_{\mathrm{L}})$ for any $x'_{\mu}$ or $y'_{\mu}$ is understood.  Here, $x_{\mu}$ represents the loop coordinate, which, along with the functional integration of the first line of (\ref{Eq:14}), will be performed shortly.  Notice that various factors of $2$, $-i$, and $g$, have been combined to produce the coefficient $+g^{2}$ multiplying (\ref{Eq:15}), and that this translated approximation to $\mathbf{L}[A]$ has become the essence of the desired eikonal function, at least before the needed Halpern integrations are performed.

	It will be most convenient to choose the zero of the coordinates $y_{\mathrm{I},0}$ and $y_{\mathrm{I\!I},0}$ at the instant of their CM distance of closest approach, which then corresponds to $s_{1} = s_{2} = 0$.  The argument of each inverse $(f \cdot \chi)$ is then independent of proper time; consider, \emph{e.g.}, the combination $(y'_{\mathrm{I}} - s_{1} \frac{p_{\mathrm{I}}}{m})_{\mu} = (y_{\mathrm{I},0}- s_{1} \frac{E}{m}; y'_{\mathrm{I},\perp}- s_{1} \frac{p_{\perp}}{m}, y_{\mathrm{I},\mathrm{L}} - s_{1} \frac{p_{\mathrm{L}}}{m})$.  But in the CM, $y_{\mathrm{I},0} = s_{1} \frac{E}{m}$, $p_{\perp} \simeq 0$, and for large momenta $y_{\mathrm{L}} \sim y_{0}$ as $p_{\mathrm{L}} \sim E$, so that $(y'_{\mathrm{I}} - s_{1} \frac{p_{\mathrm{I}}}{m})_{\mu}$ reduces to $(0; y'_{\mathrm{I},\perp},0)$.  The same argument, with the CM signs of $p_{\mathrm{I}, \mathrm{L}}$ and $y_{\mathrm{I}, \mathrm{L}}$ reversed, holds for the combination $(y'_{\mathrm{I\!I}} - s_{2} \frac{p_{\mathrm{I\!I}}}{m})_{\mu} \rightarrow (0; y'_{\mathrm{I\!I}, \perp},0) $.

	This now allows the $s_{1,2}$ integrals to be performed, and for this a Fourier representation of the two delta functions of (\ref{Eq:15}) is convenient, which yields
\begin{eqnarray}
& & m^{2} \int{\frac{\mathrm{d}^{4}k}{(2\pi)^{4}} \,  \int{\frac{\mathrm{d}^{4}q}{(2\pi)^{4}} \, (2\pi)^{2} \, \delta(q \cdot p_{\mathrm{I}}) \, \delta(k \cdot p_{\mathrm{I\!I}}) }} \\ \nonumber & & \quad \cdot e^{iq \cdot (y'_{\mathrm{I}} - x') + i k \cdot (y'_{\mathrm{I\!I}} - x'') } \cdot e^{i q \cdot v(t_{1}) + i k \cdot v(t_{2})},
\end{eqnarray}

\noindent and where $\delta(q \cdot p_{\mathrm{I}}) = \frac{1}{E} \, \delta(q_{0} - q_{\mathrm{L}} \frac{p_{\mathrm{L}}}{E}) \simeq \frac{1}{E} \, \delta(q_{0} - q_{\mathrm{L}})$, $\delta(k \cdot p_{\mathrm{I\!I}}) = \frac{1}{E} \, \delta(k_{0} + k_{\mathrm{L}} \frac{p_{\mathrm{L}}}{E}) \simeq \frac{1}{E} \, \delta(k_{0} + k_{\mathrm{L}})$, $p_{\mathrm{L}}^{} = p_{\mathrm{I} \mathrm{L}}^{} = - p_{\mathrm{I\!I} \mathrm{L}}^{}$, and $E = E_{\mathrm{I}} = E_{\mathrm{I\!I}}$.  Then, the integrations $\int{\mathrm{d}x_{0}} \int{\mathrm{d}x_{\mathrm{L}}}$ produce the additional factors $(2\pi)^{2} \, \delta(q_{0} + k_{0}) \, \delta(q_{\mathrm{L}} + k_{\mathrm{L}})$, which multiply the previous line, and produce a net combination of $\frac{1}{2} \, \frac{(2\pi)^{2}}{E^{2}} \, \delta(q_{0}) \, \delta(k_{0})\, \delta(q_{\mathrm{L}}) \, \delta(k_{\mathrm{L}})$, so that the remaining $q$- and $k$- integrals refer to transverse components only.

	Before performing the transverse integrations, it will be convenient to make one further simplification, one which appears as a reasonable approximation, but can be justified following the argument of Appendix B of Ref.~\cite{Fried2011_QCD3}.  This simplification replaces the arguments of each inverse $(f \cdot \chi)$ factor by their "expected" values $y_{\mathrm{I}\perp}$ and $y_{\mathrm{I\!I}\perp}$.  This step would appear to be a reasonable approximation because the $\mathfrak{a}(y_{\mathrm{I}\perp} - y'_{\mathrm{I}\perp})$ and $\mathfrak{a}(y_{\mathrm{I\!I}\perp} - y'_{\mathrm{I\!I}\perp})$ distributions are each peaked about a zero value of their arguments, which essentially forces the primed transverse $y'_{\perp}$ coordinates to lie close to their unprimed values.  But if appropriate care is taken in the evaluation of the $\chi$-integrations, one eventually finds the same form of result as when this simplification is first performed.  Hence, in the interest of clarity and simplicity, we now adopt the replacements:  $\chi(y'_{\perp}) \rightarrow \chi(y_{\perp})$.

	We next evaluate the multiple transverse integrals of (\ref{Eq:15}) by first writing Fourier transforms for each of the $\mathfrak{a}(z_{\perp})$ distributions,
\begin{equation}
\mathfrak{a}(z_{\perp}) = \int{\frac{\mathrm{d}^{2}\kappa}{(2\pi)^{2}} \, \tilde{\mathfrak{a}}(\kappa) \, e^{i\kappa \cdot z_{\perp}} },
\end{equation}

\noindent so that
\begin{eqnarray}\label{Eq:16}
& & (2\pi)^{-2} \, \int{\mathrm{d}^{2}q} \,  \int{\mathrm{d}^{2}k} \, \int{\mathrm{d}^{2}x'_{\perp}} \, \int{\mathrm{d}^{2}x''_{\perp} } \, \int{\mathrm{d}^{2}y'_{\mathrm{I},\perp}} \, \int{\mathrm{d}^{2}y'_{\mathrm{I\!I},\perp} } \\ \nonumber & & \quad \cdot \mathfrak{a}(x_{\perp} - x'_{\perp}) \, \mathfrak{a}(x_{\perp} - x''_{\perp}) \, \mathfrak{a}(y_{\mathrm{I}\perp} - y'_{\mathrm{I}\perp}) \, \mathfrak{a}(y_{\mathrm{I\!I} \perp} - y'_{\mathrm{I\!I} \perp})  \\ \nonumber & & \quad \quad \cdot e^{iq \cdot (y'_{\mathrm{I}\perp} - x'_{\perp}) + i k \cdot (y'_{\mathrm{I\!I}\perp} - x''_{\perp}) } \\ \nonumber & & = (2\pi)^{-2} \int{\mathrm{d}^{2}q \, |\tilde{\mathfrak{a}}(q)|^{2} \, |\tilde{\mathfrak{a}}(q)|^{2} \, e^{i q \cdot B} } \\ \nonumber & & = \int{\mathrm{d}^{2}b \, \varphi(b) \, \varphi(\vec{B} - \vec{b})} \equiv \bar{\varphi}(B),
\end{eqnarray}

\noindent where $\bar{\varphi}(B)$ will provide a slower fall-off with increasing $B$ than does $\varphi(b)$ with increasing $b$. Here, $B = y_{\mathrm{I}\perp} - y_{\mathrm{I\!I}\perp}$, and $\varphi(b)$ is the modified statement of transverse imprecision introduced in \cite{Fried2009_QCD1} and made precise in \cite{Fried2011_QCD3}, $\varphi(b) \simeq \frac{\mu^{2}}{\pi} \, e^{-(\mu b)^{2+\xi}}$, $\xi \ll 1$.

	The integral of (\ref{Eq:16}) is not the final statement of $B$ dependence, because a term proportional to $q_{\alpha} \, q_{\beta}$ arising from the evaluation of the functional integral of (\ref{Eq:14}) and appearing in (\ref{Eq:20}) must still be included.  One requires
\begin{eqnarray}\label{Eq:17}
\mathcal{N} \int{\mathrm{d}[v] \, e^{\frac{i}{2} \int{v \cdot (2h)^{-1} \cdot v}} \, v'_{\alpha}(t_{1}) \, v'_{\beta}(t_{2}) \, \delta^{(4)}(v(t)) \, e^{i q \cdot (v(t_{1}) - v(t_{2}))}   },
\end{eqnarray}

\noindent which may be accomplished by inserting a Fourier representation of $\delta^{(4)}(v(t))$, and rewriting (\ref{Eq:17}) as
\begin{eqnarray}\label{Eq:18}
& & \frac{1}{i} \, \frac{\partial}{\partial t_{a}} \, \frac{\delta}{\delta g_{\alpha}(t_{a})} \, \frac{1}{i} \, \frac{\partial}{\partial t_{b}} \, \frac{\delta}{\delta g_{\beta}(t_{b})} \, \\ \nonumber & & \quad \cdot \left. \mathcal{N} \,  \int{\mathrm{d}[v] \, e^{\frac{i}{2} \int{v \cdot (2h)^{-1} \cdot v} + i \int_{0}^{t}{\mathrm{d}t' \, v_{\mu}(t') [f_{\mu}(t') + g_{\mu}(t')]  } }} \right|_{t_{a} \rightarrow t_{1}, t_{b} \rightarrow t_{2} },
\end{eqnarray}

\noindent where $f_{\mu}(t') = p_{\mu} \, \delta(t' - t)$, and $g_{\mu}(t') = q_{\mu} \, [\delta(t' - t_{1}) - \delta(t' - t_{2})]$. The normalized, Gaussian functional integral of (\ref{Eq:18}) is then
\begin{eqnarray}\label{Eq:19}
\exp{\left\{ -i \int_{0}^{t}{\mathrm{d}t' \, \int_{0}^{t}{\mathrm{d}t'' \, \left[f_{\mu}(t') + g_{\mu}(t') \right] \, h(t',t'') \, [f_{\mu}(t'') + g_{\mu}(t'')] }} \right\} },
\end{eqnarray}

\noindent and the functional and conventional derivatives of (\ref{Eq:18}), as well as the resulting Gaussian $\int{\mathrm{d}^{4}p}$ are immediate.  Combining all factors, one obtains for the translated and simplified $\mathbf{L}[A]$ of (\ref{Eq:14}) the result
\begin{eqnarray}\label{Eq:20}
& & \frac{i}{4} \, \left( \frac{p_{\mathrm{I}\mu} \, p_{\mathrm{I\!I}\nu} }{E^{2}} \right) \, \frac{g^{2}}{(4\pi)^{2}} \, \int{\mathrm{d}^{2}q \, |\tilde{\mathfrak{a}}(q)|^{2} \, |\tilde{\mathfrak{a}}(q)|^{2} \, e^{i q \cdot B} } \, \int_{0}^{\infty}{\frac{\mathrm{d}t}{t} \, e^{-itm^{2}} \, \bar{\Omega}^{a} \, \bar{\Omega}^{b} } \\ \nonumber & & \quad \cdot \mathcal{N}' \, \int{\mathrm{d}[\hat{\alpha}] \,  \int{\mathrm{d}[\hat{\Omega}] \, e^{i \int_{0}^{t}{\mathrm{d}t' \, \hat{\alpha}^{a}(t') \, \hat{\Omega}^{a}(t')}}  }} \, \Tr{\left( e^{i \int_{0}^{t}{\mathrm{d}t' \hat{\alpha}^{a}(t') \, \lambda^{a}}   } \right)_{+}} \\ \nonumber && \quad \cdot \int_{0}^{1}{\mathrm{d}\mathfrak{z}_{1} \, \int_{0}^{1}{\mathrm{d}\mathfrak{z}_{2} \, \left\{ \frac{2i}{t} \, \delta_{\alpha \beta} \, [\delta(\mathfrak{z}_{1} - \mathfrak{z}_{2} ) -1 ] + q_{\alpha} q_{\beta}[1 - 4 (\mathfrak{z}_{1} - \mathfrak{z}_{2})^{2}]\right\} \, \hat{\Omega}^{c}(\mathfrak{z}_{1} \, t) \, \hat{\Omega}^{d}(\mathfrak{z}_{2} \, t) } } \\ \nonumber & & \quad \quad \cdot \left. \left( f \cdot \chi(y_{\mathrm{I} \perp}) \right)^{-1} \right|_{\mu \alpha}^{a c} \cdot \left. \left( f \cdot \chi(y_{\mathrm{I\!I} \perp}) \right)^{-1} \right|_{\beta \nu}^{d b},
\end{eqnarray}

\noindent where we have replaced $t_{1,2}$ by $t \cdot \mathfrak{z}_{1,2}$, and $\int_{0}^{t}{\mathrm{d}t_{1,2}}$ by $t \int_{0}^{1}{\mathrm{d}\mathfrak{z}_{1,2}}$         .

	Eq.~(\ref{Eq:20}) is noteworthy for several reasons, among which is the special way in which the manifest gauge invariance of $\mathbf{L}[A]$ is displayed in the automatic cancelation of the quadratic divergence associated with the removal of the $\int_{0}{\frac{\mathrm{d}t}{t^{2}}}$ of (\ref{Eq:20}).  In Feynman graph language this does not happen automatically, for the divergence of the fermion loop "overpowers" the gauge invariance of the basic theory; and one must resort to other measures to remove that quadratic divergence.  As Schwinger pointed out long ago~\cite{Schwinger1951}, in his functional development of radiative corrections to QED in terms of proper time variables, such unwanted and improper terms never appear in calculations so defined.

	The gauge-invariant divergence of this loop is logarithmic, as expected; and its renormalization displays the behavior associated with the property of "anti-shielding", as expected in QCD, rather than the "shielding" of QED. This divergence, associated with the lower limit of $0$ in the $t$-integral of (\ref{Eq:20}), may be described in configuration space by replacing that lower limit by a small quantity $\epsilon$, of dimensions of $(\mbox{length})^{2}$; in momentum space, this would corresponds to a cut-off of $\Lambda^{2} = 1/\epsilon$.  It will be convenient to perform the variable change $t = \epsilon r$, and then rotate contours $r \rightarrow -i z$, so that the $t$-integral of (\ref{Eq:20}) becomes
\begin{eqnarray}\label{Eq:21}
& & \int_{1}^{\infty}{\frac{\mathrm{d}z}{z} \, e^{-\frac{z}{\Lambda^{2}}\, [m^{2} + q^{2} |\mathfrak{z}_{12}^{}| \, (1 - |\mathfrak{z}_{12}^{}|)] } \,  \hat{\Omega}^{c}(\frac{-iz}{\Lambda^{2}} \, \mathfrak{z}_{1}) \, \hat{\Omega}^{d}(\frac{-iz}{\Lambda^{2}} \, \mathfrak{z}_{2}) } \\ \nonumber & & \simeq \left\{ \ln{\left( \frac{\Lambda^{2}}{m^{2}} \right)} - \ln{\left[ 1 + \frac{q^{2}}{m^{2}} \, |\mathfrak{z}_{12}^{}| \, (1 - |\mathfrak{z}_{12}^{}|) \right]} \right\} \, \hat{\Omega}^{c}(0) \, \hat{\Omega}^{d}(0),
\end{eqnarray}

\noindent where $\mathfrak{z}_{12} = \mathfrak{z}_{1} - \mathfrak{z}_{2}$, and we have allowed $\Lambda$ to become arbitrarily large in the arguments of $\hat{\Omega}^{c}$ and $\hat{\Omega}^{d}$; we have also, for the moment, suppressed the $t$-dependent integrals coupling $\hat{\alpha}^{a}$ and $\hat{\Omega}^{b}$ to the Gell-Mann matrices $\lambda^{c}$.  The renormalized coupling of this order $g^{2}$ bundle diagram may be defined by the relation suggested by (\ref{Eq:21}), as
\begin{eqnarray}\label{Eq:22}
g_{\mathrm{R}}^{2}(q^{2}) = g^{2} \,  \ln{\left[ \frac{\Lambda^{2}} {m^{2} + q^{2} \, |\mathfrak{z}_{12}^{}| \, (1 - |\mathfrak{z}_{12}^{}|)} \right]}
\end{eqnarray}

\noindent which displays the expected QCD form, of an effective, or (partially) renormalized coupling that decreases with increasing momentum transfer.  And since the $q$-values expected from its subsequent integration are less than the quark mass, and both are understood to be far less than any realistic cut-off adopted for $\Lambda$, (\ref{Eq:22}) may be most simply approximated by
\begin{eqnarray}\label{Eq:23}
g_{\mathrm{R}}^{2} = g^{2} \,  \ln{\left( \frac{\Lambda^{2}}{m^{2}} \right)},
\end{eqnarray}
					
\noindent where it is clear that the bare coupling $g$ of the original Lagrangian is smaller than the renormalized coupling, in contrast to Abelian QED, where the reverse holds.

	From this example one sees that our formalism is non-perturbative in the sense of summing over all gluon exchanges between specified quarks; but that if one of those quark lines is part of a closed loop, then a perturbation expansion can be defined involving increasing numbers of gluon bundles exchanged between that closed loop and other, specified quarks, which may themselves be associated with other quark loops.  Can the non-perturbative nature of our analysis be extended to include all possible $\mathbf{L}[A]$ interactions?  We hope to answer this very non-trivial question in a subsequent publication.

	The color dependence of (\ref{Eq:15}) remains to be treated, and for this it is simplest to return to that stage of calculation before renormalization was discussed.  There, the factors of $\hat{\Omega}^{c}(t_{1})$ and $\hat{\Omega}^{d}(t_{2})$ remain to be evaluated, which process consists of converting them into Gell-Mann matrices $\lambda^{c}$ and $\lambda^{d}$.  %For continuity of presentation, that replacement has been derived in both a formal and a detailed manner in the Appendix; and the result of either argument is
It can easily be shown that the commuting factors of $\hat{\Omega}^{c}(t_{1}) \, \hat{\Omega}^{d}(t_{2})$ are to be replaced by $\lambda^{c} \lambda^{d} \theta(t_{1} - t_{2}) + \lambda^{d} \lambda^{c} \theta(t_{2} - t_{1})$, while, simultaneously, the functional integrations over $\hat{\alpha}$ and $\hat{\Omega}$ have diappeared.
%$\int{\mathrm{d}[\hat{\alpha}]}$ and $\int{\mathrm{d}[\hat{\Omega}]}$ have disappeared.

	After renormalization, in which the $\int{\mathrm{d}t}$ is effectively evaluated close to its lower limit, and where $t_{1,2} \Rightarrow t \cdot \mathfrak{z}_{1,2}$, as $t \rightarrow 0$, $\theta(t_{1} - t_{2}) \rightarrow \theta(t_{2} - t_{1}) \rightarrow \theta(0) = 1/2$, and the product $\hat{\Omega}^{c}(t_{1}) \, \hat{\Omega}^{d}(t_{2})$ is replaced by $\frac{1}{2} \, \{ \lambda^{c}, \lambda^{d} \}$.  As noted above, for simplicity and ease of presentation, we have neglected quark spin dependence, and its associated $\lambda$-dependence, so that the $\Tr{}$ operation over both Dirac and color indices yields $\Tr{[\frac{1}{2} \, \{ \lambda^{c}, \lambda^{d} \}]} = 8\, \delta^{cd}$.  Eq.~(\ref{Eq:20}) then reduces to			
\begin{eqnarray}
& & \left( \frac{p_{\mathrm{I}\mu} \, p_{\mathrm{I\!I}\nu} }{E^{2}} \right) \, \frac{g_{\mathrm{R}}^{2}}{3\pi^{2}} \, \bar{\Omega}^{a} \, \bar{\Omega}^{b} \, \left. \left( f \cdot \chi(y_{\mathrm{I} \perp}) \right)^{-1} \right|_{\mu \alpha}^{a c} \cdot \left. \left( f \cdot \chi(y_{\mathrm{I\!I} \perp}) \right)^{-1} \right|_{\beta \nu}^{c b} \\ \nonumber & & \quad \cdot \left( - \frac{\partial}{\partial B_{\alpha}} \, \frac{\partial}{\partial B_{\beta}} \right) \, \bar{\varphi}(B),
\end{eqnarray}

\noindent where the $q_{\alpha} q_{\beta}$ factors of (\ref{Eq:20}) have been replaced by $\left( - \frac{\partial}{\partial B_{\alpha}} \, \frac{\partial}{\partial B_{\beta}} \right)$.

	The relevant space-time indices enter here in the form
\begin{eqnarray}\label{Eq:24}
\left( \frac{p_{\mathrm{I}\mu} \, p_{\mathrm{I\!I}\nu} }{E^{2}} \right) \, \left. \left( f \cdot \chi(\mathrm{I}) \right)^{-1} \right|_{\mu \alpha}^{a c} \cdot \left. \left( f \cdot \chi(\mathrm{I\!I}) \right)^{-1} \right|_{\beta \nu}^{c b},
\end{eqnarray}

\noindent and, remembering the antisymmetry of each element's color and space-time indices, and that the $\alpha$, $\beta$ are transverse indices, Eq.~(\ref{Eq:24}) may be rewritten as
\begin{eqnarray}\label{Eq:24b}
+i \left. \left( f \cdot \chi(\mathrm{I}) \right)^{-1} \right|_{4 \alpha}^{a c} \cdot \left. \left( f \cdot \chi(\mathrm{I\!I}) \right)^{-1} \right|_{\beta \mathrm{L}}^{b c} - i \left. \left( f \cdot \chi(\mathrm{I}) \right)^{-1} \right|_{\mathrm{L} \alpha}^{a c} \cdot \left. \left( f \cdot \chi(\mathrm{I\!I}) \right)^{-1} \right|_{4 \beta}^{b c},
\end{eqnarray}

\noindent  where, because the longitudinal and energy components are far larger than the transverse momenta, the $\mu$, $\nu$ indices correspond to $0$ and $\mathrm{L}$ only.  Using the Minkowski metric, where $\chi_{4 \alpha}^{a} \equiv i \chi_{0 \alpha}^{a}$, then $\left. (f \cdot \chi)^{-1} \right|_{4 \alpha} = -i 	\left. (f \cdot \chi)^{-1} \right|_{0 \alpha}$.  Further, in the CM system, where the longitudinal projections of $y_{\mathrm{I}}$ and $y_{\mathrm{I\!I}}$ point in exactly opposite directions, while the $\chi$ variables depend only upon their respective transverse coordinates; then, the CM longitudinal projections of such $\chi^{-1}$ will point in opposite directions.  In order to have similar, if arbitrary, constructions of the $\chi(\mathrm{I})$ and $\chi(\mathrm{I\!I})$, we set $\chi'_{\beta \mathrm{L}}(\mathrm{I\!I}) = - \chi(\mathrm{I\!I})|_{\beta \mathrm{L}}$, so as to bring (\ref{Eq:24b}) into the form
\begin{eqnarray}\label{Eq:25}
- \left\{ \left. \left( f \cdot \chi(\mathrm{I}) \right)^{-1} \right|_{\alpha 0}^{c a} \cdot \left. \left( f \cdot \chi'(\mathrm{I\!I}) \right)^{-1} \right|_{\beta \mathrm{L}}^{c b} + \left. \left( f \cdot \chi(\mathrm{I}) \right)^{-1} \right|_{\alpha \mathrm{L}}^{c a} \cdot \left. \left( f \cdot \chi(\mathrm{I\!I}) \right)^{-1} \right|_{\beta 0}^{c b} \right\}.
\end{eqnarray}

\noindent The normalized integrals over $\int{\mathrm{d}^{n}\chi_{\beta \mathrm{L}}(\mathrm{I\!I})}$ and $\int{\mathrm{d}^{n}\chi'_{\beta \mathrm{L}}(\mathrm{I\!I})}$, are the same, and are unchanged; and since the values of $y_{\mathrm{I} \perp}$ and $y_{\mathrm{I\!I} \perp}$ appearing in the arguments of each $\chi$ serve only to indicate that two separate integrations are required, one can interchange those arguments in the second term of (\ref{Eq:25}) to obtain, in place of (\ref{Eq:25}),
\begin{eqnarray}\label{Eq:26}
- \left\{ \left. \left( f \cdot \chi(\mathrm{I}) \right)^{-1} \right|_{\alpha 0}^{c a} \cdot \left. \left( f \cdot \chi(\mathrm{I\!I}) \right)^{-1} \right|_{\beta \mathrm{L}}^{c b} + \left. \left( f \cdot \chi(\mathrm{I}) \right)^{-1} \right|_{\beta 0}^{c b} \cdot \left. \left( f \cdot \chi(\mathrm{I\!I}) \right)^{-1} \right|_{\alpha \mathrm{L}}^{c a} \right\},
\end{eqnarray}

\noindent a result which is explicitly symmetric in $a$ and $b$ and in $\alpha$ and $\beta$.

	As in previous discussion of Refs.~\cite{Fried2009_QCD1} and \cite{Fried2011_QCD4}, we assume that each $\chi^{c}$ can be represented by an angular projection $\mathfrak{z}^{c}$ multiplying a magnitude $R$, $\chi^{c} = \mathfrak{z}^{c} \, R$, and we now suppress the result of those normalized angular integrals, assuming that the most significant behavior of our results is due to integration over the magnitudes.  Of course, such a simplification must be checked by detailed, numerical calculation; but this would appear to be a reasonable approximation.  Note that the index symmetries of (\ref{Eq:25}) would be enforced by multiplication by $q_{\alpha} q_{\beta}$ of (\ref{Eq:20}), and by the $\bar{\Omega}^{a} \bar{\Omega}^{b}$, corresponding to color singlet gluon emission and absorption of the two nucleons.  There is then no difference between the two terms of (\ref{Eq:26}); they are both going to give the same contribution, and so (\ref{Eq:26}), after multiplication by the $\bar{\Omega}^{a,b}$, and $q_{\alpha}, q_{\beta}$ is equivalent to
\begin{eqnarray}\label{Eq:27}
-2 \, \bar{\Omega}^{a} \bar{\Omega}^{b} \, q_{\alpha} q_{\beta} \,  \left. \left( f \cdot \chi(\mathrm{I}) \right)^{-1} \right|_{0 \alpha}^{a c} \cdot \left. \left( f \cdot \chi(\mathrm{I\!I}) \right)^{-1} \right|_{0 \beta}^{b c}.
\end{eqnarray}												

	The attentive reader will notice that there is one aspect of our procedure of obtaining an effective potential from an eikonal function which remains to be discussed: what is to be done when the eikonal itself contains transverse components of coordinates or corresponding momentum transfer?  Physically, each component of the initial  momentum transfer $q_{\mu}$ of nucleon $\mathrm{I}$ must be transferred to the corresponding component of momentum transfer of nucleon $\mathrm{I\!I}$ on the other side of the loop, $q_{1}(\mathrm{I})$ to $q_{1}(\mathrm{I\!I})$ and $q_{2}(\mathrm{I})$ to $q_{2}(\mathrm{I\!I})$; in other words, a $\delta_{\alpha \beta}$ must appear in (\ref{Eq:27}), either from integrations over the "angular" components of the Halpern variables, which we have suppressed, or as a definite statement of our procedure, which we now state: All such "free" indices are to be averaged over, a stipulation which has consequences in other contexts (renormalization theory and nuclear binding).  In the present case, it means that $q_{\alpha} q_{\beta}$ is to be replaced by $ \frac{1}{2} q^{2} \delta_{\alpha \beta}$ as is physically necessary. Then, we may write the simplified, normalized integrals to be performed as
\begin{eqnarray}\label{Eq:28}
\mathcal{N} \, \int_{0}^{\infty}{\mathrm{d}R_{\mathrm{I}} \, R_{\mathrm{I}}^{3} \, \int_{0}^{\infty}{\mathrm{d}R_{\mathrm{I\!I}} \, R_{\mathrm{I\!I}}^{3} \, e^{\frac{i}{4}(R_{\mathrm{I}}^{2} + R_{\mathrm{I\!I}}^{2}) - i\frac{\mathcal{C}(B,E)}{R_{\mathrm{I}} R_{\mathrm{I\!I}}}  } }},
\end{eqnarray}	

\noindent where $\mathcal{C}(B,E) = \frac{1}{3} \, \left( \frac{g_{\mathrm{R}}^{2}}{4 \pi}\right) \, (\delta^{2})^{2} \,  \left[ (- \nabla_{B}^{2}) \cdot \bar{\varphi}(B) \right]$, and, as explained in detail in Ref.~\cite{Fried2011_QCD3}, $\delta^{2}$ is the scale change needed when passing from the Halpern FI to the individual $\int{\mathrm{d}^{8}\chi}$: $\delta^{2} = (\mu E)^{-1}$.  The $R_{\mathrm{I},\mathrm{I\!I}}^{3}$, rather than the $R_{\mathrm{I},\mathrm{I\!I}}^{7}$, result from a factor $\left[ R_{\mathrm{I},\mathrm{I\!I}}^{8} \right]^{-1/2}$ extracted from each determinantal factor $[\det{(g f \cdot \chi)}]^{\frac{1}{2}}$ of (\ref{Eq:11}).

	It is the double derivatives with respect to $B$, the impact parameter between the two nucleons, arising from the $q^{2} \delta_{\alpha \beta}$ components of the closed-loop integral, which provides the sign of a potential that produces nucleon binding; and it is in this qualitative possibility of generating a "model deuteron" from two bound nucleons that the possibility of obtaining true Nuclear Physics from transversally averaged QCD appears.

\section{\label{SEC3}A Qualitative Binding Potential}

Before passing to the final steps of the calculation of this potential, we remind the reader that this treatment is based on the simplest possible realization of realistic QCD, based on a single, massive quark interacting with its complement of SU(3) massless gluons; flavors and electroweak interactions, as well as quark and nucleon spins and angular momenta have been neglected, for simplicity, and can be added separately, producing definite variations of the potential below.  The rigorous property of Effective Locality, defined and discussed in detail in Refs.~\cite{Fried2010_QCD2}, \cite{Fried2011_QCD3} and \cite{Fried2011_QCD4}, immensely simplifies the original Halpern FI of Ref.~\cite{Fried2009_QCD1} by reducing it, in the present case, to two sets of ordinary integrals; and we have here suppressed the "angular" color integrations, retaining dependence only on the magnitudes of the reduced Halpern variables, an approximation which must be verified by numerical calculation. Nevertheless, it should be of more than passing interest to see just how a qualitatively reasonable nucleon potential can appear from such basic QCD.

Of course, that potential is not meant to suggest that two neutrons will bind, for their fermionic nature has been suppressed with the neglect of their spins; nor would it be suggestive of two protons binding to form a nucleus, because both spin structure and electrodynamics have been omitted.  That potential is not yet meant to be compared with precise experimental data, except in the sense of its qualitative behavior, producing for two distinguishable nucleons scattering at high relative energies, as well as the possibility of binding into a "model deuteron" at lower incident energies.

	With the simplifications and approximations discussed in the preceding Sections, we now write (\ref{Eq:7}) in the form
\begin{eqnarray}\label{Eq:29}
e^{i \mathbb{X}(B,E)} = \mathcal{N} \, \int_{0}^{\infty}{\mathrm{d}R_{\mathrm{I}} \, R_{\mathrm{I}}^{3} \, \int_{0}^{\infty}{\mathrm{d}R_{\mathrm{I\!I}} \, R_{\mathrm{I\!I}}^{3} \, e^{\frac{i}{4}(R_{\mathrm{I}}^{2} + R_{\mathrm{I\!I}}^{2}) - i\frac{\mathcal{C}(B,E)}{R_{\mathrm{I}} R_{\mathrm{I\!I}}}  } }},
\end{eqnarray}	

\noindent where we make the further, simplifying approximation of suppressing the parameter $\xi \simeq 0.1$, of $\varphi(b)$, which was crucial in the construction of quark binding, but would here only slightly change the shape of the nucleon binding potential. Setting then $\varphi(b) = (\mu^{2}/\pi) \, \exp{[-(\mu b)^{2}]}$, one finds
\begin{equation}
\mathcal{C}(B,E) = \frac{g_{\mathrm{R}}^{2}}{6 \pi^{2}} \left( \frac{\mu}{E} \right)^{2} \, \left[ 2 - \mu^{2} B^{2} \right] \, e^{-\frac{\mu^{2} B^{2}}{2}}.
\end{equation}

	There are several methods of obvious approximation to the integral of (\ref{Eq:29}):
\begin{enumerate}
  \item\label{item:1}  A change of variables to polar coordinates, $R_{\mathrm{I}} = R \sin(\theta)$, $R_{\mathrm{I\!I}} = R \cos(\theta)$, for which the radial integral can be done immediately, but the subsequent angular integral requires an approximation.

  \item\label{item:2} Both $R_{\mathrm{I},\mathrm{I\!I}}$ integrands correspond to a function rising as $|R_{\mathrm{I},\mathrm{I\!I}}|$ increases from zero, and then falling away to $0$ as these coordinates become large; and both may be approximated by (different) Gaussian approximations.
  \item\label{item:3} Both (\ref{item:1}) and (\ref{item:2}) lead to rather complicated expressions involving fractional powers of complex functions.  There is, however, a simpler approximation, available in this eikonal context where $\mathcal{C}(B,E)$ contains the factor $\delta^{4} \mu^{4} = (\mu / E)^{2}$, where $E$ is the CM energy of the scattering nucleons.  In conventional eikonal representations, there is always a dimensionless, energy-dependent, kinematical factor, $\gamma(E)$, multiplying an impact-parameter-dependent function which is the "true" eikonal function, derived from an initial potential function; and if that combination is small, then the final eikonal amplitude may have its exponential factor expanded, so that only the linear dependence of that exponential is retained. Here, that energy-dependent $\delta^{4} \mu^{4}$-factor is surely small, but is it the correct $\gamma(E)$?  In Potential Theory and in various forms of QFT, the functional form of $\gamma(E)$ can vary widely, but we have no precedent here to specify the "correct" form of $\gamma(E)$ to appear upon the exchange of a pair of gluon bundles supporting a quark loop.  We shall therefore make the simplest choice of adopting $[\mu \delta(E)]^{4}$ as our tentative $\gamma(E)$; and at the very end of the calculation return to see if this choice is consistent with the order-of-magnitude of our qualitative potential.
\end{enumerate}

	We now expand to first order both the eikonal amplitude of (\ref{Eq:7}), which is the left-hand-side of (\ref{Eq:29}), and the exponential factor containing $\mathcal{C}(B,E)$ of (\ref{Eq:29}), so that
\begin{eqnarray}
i \mathbb{X}(B,E) &=& i\mathbb{X} \gamma(E) = -i \mathcal{C}(B,E) \, \mathcal{N} J^{2}, \\ \nonumber & & J = \int_{0}^{\infty}{\mathrm{d}R \, R^{3} \, e^{\frac{i}{4} R^{2}}},\\ \nonumber & & \mathcal{N} J^{2} = - i \frac{\pi}{4},
\end{eqnarray}

\noindent where the "true" eikonal function is
\begin{equation}\label{Eq:30}
\mathbb{X} \simeq \left(\frac{i}{6}\right) \, \left(\frac{g_{\mathrm{R}}^{2}}{4 \pi}\right) \, \left[ 2 - \mu^{2} B^{2} \right] \, e^{-\frac{\mu^{2} B^{2}}{2}}.
\end{equation}

\noindent This situation differs from that of the quark binding calculation of Ref.~\cite{Fried2011_QCD4}, where the large impact parameter of interest generated a large eikonal function, but a small amplitude; here, both the eikonal and the amplitude are small.

	The relation between the eikonal function and the effective potential is
\begin{equation}
\mathbb{X} = - \int_{-\infty}^{+\infty}{\mathrm{d}z_{\mathrm{L}} \, V(\vec{B} + \hat{\mathbf{P}}_{\mathrm{L}} z_{\mathrm{L}}) }
\end{equation}

\noindent and as explained in Ref.~\cite{Fried2011_QCD4}, the eikonal is real for a purely scattering potential, $V = V_{\mathrm{S}}$; but for a potential which can lead to binding, or to the production of other particles, the potential chosen must have the form $V = V_{\mathrm{S}} - i V_{\mathrm{B}}$, so that the eikonal which corresponds to binding is imaginary.  The reason is unitarity, since if extra particles, or a new bound state can be produced, the amplitude of the initial state must be reduced.  In our reversed situation, starting from the construction of a QCD amplitude, we find a clear signal of a binding potential, with $V_{\mathrm{B}}$ appearing as a real quantity,
\begin{equation}
\int_{-\infty}^{+\infty}{\mathrm{d}z_{\mathrm{L}} \, V_{\mathrm{B}} } = \frac{1}{12} \, \frac{g_{\mathrm{R}}^{2}}{4 \pi} \, \left( 2 - \mu^{2} B^{2} \right) \, e^{-\mu^{2} B^{2}/2}.
\end{equation}

	To obtain the effective potential one first calculates the two-dimensional Fourier transform of $-i \mathbb{X}(B)$, which can be expressed as proportional to
\begin{equation}\label{Eq:31}
\left( \frac{\mu^{2}}{4 \pi} \right) \, k_{\perp}^{2} \, \int{\mathrm{d}^{2}B \, e^{i k_{\perp} \cdot B - \mu^{2} B^{2}/2}},
\end{equation}

\noindent then continue $k_{\perp}^{2}$ to three dimensions, $k_{\perp}^{2} \rightarrow k_{\perp}^{2} + k_{\mathrm{L}}^{2}$, and calculate the three-dimensional Fourier transform of (\ref{Eq:31}), which yields, after removing the factor $\gamma(E) = (\mu \delta)^{4}$,
\begin{equation}\label{Eq:32}
V(r) \simeq c \, \left(\frac{g_{\mathrm{R}}^{2}}{4 \pi}\right) \, \mu \left[ 2 - \mu^{2} r^{2} \right] \, e^{-\frac{\mu^{2} r^{2}}{2}},
\end{equation}

\noindent with $c = \frac{1}{6} \, (2\pi)^{-3/2}$.  At high energies and large momentum transfers, this potential when multiplied by $(-i)$ corresponds to an effective scattering potential.

	The form of this potential is sketched in Fig.~\ref{Fig:3}, and it will look familiar to those who have inferred a nucleon potential from experimental scattering data, starting with the potentials of the 1951 paper of Jastrow~\cite{Jastrow1951}.  It must be noted that this potential is not meant to be relevant at distances  $\mu r < 1$, which is where the multiple gluon exchanges of the gluon bundles of Fig.~\ref{Fig:2}, as well as those of the omitted gluon-binding interactions of each triad of bound nucleons take place.  And of course, we have neglected electromagnetic effects, as well as all spin and angular momentum modifications, which can be included in more detailed estimations.

We have two parameters at our disposal, the mass scale $\mu  \simeq  m_{\pi}$, and  $g_{\mathrm{R}}^{2}/4\pi$, which can be chosen so as to produce a ground state with a binding energy of -2.2 MeV.  Of course, from the crudeness of the approximations made in our various estimations, we would be happy to obtain a binding energy to within a factor of 10 of this numerical result, but as it happens, we shall do somewhat better.  The corresponding calculation is demonstrated in the next Section, using the elementary Quantic technique~\cite{Quantics JMLL} of estimating a ground state.  But, simplifying approximations aside, this is clearly a potential which can bind a pair of distinct, uncharged nucleons; and it is obtained analytically, from basic transversally-averaged QCD.

%
%***********************************************************
%***********************************************************
%

%
% need to use 'JPG-PNGtoEPS' converter by Richard Socher for batch conversion via 'Sam2p' by Szabó Péter
%

\begin{figure}
\includegraphics[height=35mm]{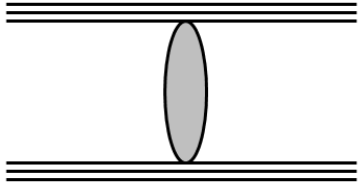}%
\caption{\label{Fig:1}A gluon bundle exchanged between two nucleons}
\end{figure}

\begin{figure}
\includegraphics[height=35mm]{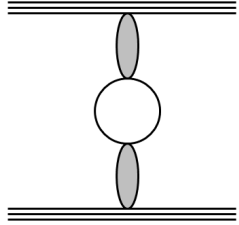}%
\caption{\label{Fig:2}Quark loop exchange through gluon bundles between two nucleons}
\end{figure}

\begin{figure}
\includegraphics[height=35mm]{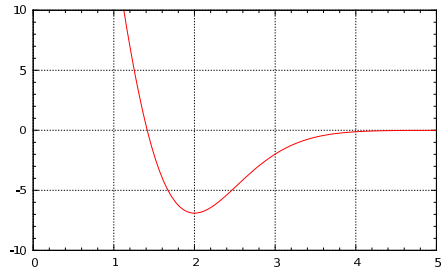}%
\caption{\label{Fig:3}Nucleon potential}
\end{figure}

%
%***********************************************************
%***********************************************************
%

%
%
%
\section{\label{SEC4}Binding Estimations}
%\subsection{}
%\subsubsection{}

The non-relativistic Hamiltonian of two, equal mass particles interacting with the above potential is
\begin{equation}\label{Eq:33}
E = \frac{p^{2}}{m} + V(r) \rightarrow \frac{1}{m r^{2}} + V(r).
\end{equation}

\noindent One can write this non-relativistic energy in dimensionless form as
\begin{equation}\label{Eq:34}
\frac{E(y)}{M} = \frac{1}{y^{2}} + \frac{V_{0}}{M} \, [ 2 - y^{2}] \, e^{-\frac{y^{2}}{2}}, \quad y = \mu r,
\end{equation}
	
\noindent where we have set $\mu = m_{\pi}$.  In units of MeV, $M = \mu^{2}/m = 18.2$, $g_{\mathrm{R}}^{2}/4\pi = \aleph \cdot 10$, and, combining all the relevant factors of the previous paragraphs, $ V_{0} = 14.3 \, \aleph$ (MeV).

	The Quantic method~\cite{Quantics JMLL} of estimating a ground state is to rewrite $p$ as $1/r  = \mu /y$, to find the minimum of $E(y)$, and use that minimum point $y_{0}$ to define $E(y_{0})$, which is to be interpreted as a qualitative estimate of the ground-state energy. The minimization statement is given by the vanishing of the derivative of (\ref{Eq:34}) at $y = y_{0}$,
\begin{equation}\label{Eq:35}
0 = - \frac{2}{y_{0}^{3}} + \frac{V_{0}}{M} \, y_{0} \, [ y_{0}^{2} - 4] \, e^{-\frac{y_{0}^{2}}{2}},
\end{equation}

\noindent and the customary way of solving such a problem is to solve (\ref{Eq:35}) for $y_{0}$, and then substitute that value of $y_{0}$ back into (\ref{Eq:34}) to obtain the binding energy.  But since $\aleph$ is essentially unknown -- one might guess it to be on the order of 1, representing a strong, nuclear force -- and because we do want to represent the bound state energy as $E(y_{0}) = -2.2$ Mev, let us use that number together with the value of $M$ to solve for $y_{0}$; and then solve for the value of $\aleph$.

	To do this, combine (\ref{Eq:34}) and (\ref{Eq:35}) in such a manner that the $\exp{[-y_{0}^{2}]}$ factors of both equations are canceled, which produces
\begin{equation}\label{Eq:36}
y_{0}^{2} \, (y_{0}^{2} - 4) \, \left[ 1 + y_{0}^{2} \frac{|E|}{M} \right]= 2 (y_{0}^{2} - 4)
\end{equation}

\noindent which is a cubic equation in $y^{2}$.  From the graph of Fig.~\ref{Fig:3}, one sees that the minimum of the potential lies close to $y_{0} = 2$, which suggests that the minimum of the energy should be somewhat larger; and this suggests the choice $y_{0} = 3 + \Delta$ as a reasonable choice for the approximate solution of (\ref{Eq:36}), retaining terms of no higher order than $\Delta$ (under the subsequently verified assumption that $\Delta^{2} \ll 6|\Delta|$).  This leads to the result: $\Delta \simeq .33$ and $y_{0} \simeq 3.33$.  Upon substituting this value of $y_{0}$ into (\ref{Eq:34}) there follows $\aleph \simeq 1.25$, which provides the expected order of magnitude for a strong-coupling process.  One may expect that when contributions from quark and nucleon spins are included, that number will decrease slightly, retaining its strong-coupling character.

%which is perhaps twice as large as expected, but can be accepted as reasonable because we are asking our equations in which spin considerations have been omitted, to provide the experimental bound-state energy, which surely includes contributions from quark and nucleon spins.

%
%
%
\section{\label{SEC5}Summary and Speculation}
%\subsection{}
%\subsubsection{}

	While the arguments put forth above are concerned with a realistic version of QCD, and have for simplicity neglected flavors, and electroweak interactions, quark and nucleon spin dependence, and have suppressed several "angular" integrations, all of which can be restored, as desired, the result is an explicit, model "deuteron" potential, of sufficiently short range and of the right order of magnitude to be considered as a qualitative derivation of nucleon-nucleon forces from basic realistic QCD.

	Our tentative choice of  $\gamma(E) = (\mu \, \delta)^{4} = (\mu/E)^{2}$ has turned out to be qualitatively correct; and, in an eikonal context, this is interesting because it suggests that for the exchange of a composite object -- in this case, the gluon bundles supporting a quark loop -- between two "scalar" nucleons, the $\gamma(E)$ factor is not just what one typically finds when exchanging scalar quanta, $(m/E)^{2}$, but retains the memory and has a signature of the composite structure being exchanged, $\gamma(E) = (\mu/m)^{2} \, (m/E)^{2}$.

	The above analysis should be almost immediately applicable to high-energy nucleon-nucleon scattering; and it will be interesting to see if the result of that calculation corresponds to the physical arguments recently suggested by Islam~\cite{Islam}.

	Generalizations of this two-nucleon deuteron model to the construction of heavier, stable nuclei may well be possible, and might provide at least a partial basis for the nuclear shell model and the independent boson (IBM) model.  In the first case, one would ask how many nucleon-generated gluon bundles can be attached to a single quark loop; and for the IBM model, asking how effective would attractive pairwise interactions of the deuteron form be when exchanged between nucleons in a three-dimensional array.

	Finally, on a more fundamental level, it will be most interesting to see just how the structure of renormalization theory turns out for realistic QCD, the theory which has, built-in, quark transverse imprecision.  From the experience gained in our work so far, the simplifications in which non-perturbative gluon exchanges organize themselves into gluon bundle exchange displaying Effective Locality suggest that truly non-perturbative renormalization will turn out to be simpler than that of QED.  We hope to answer this question in the near future.

\begin{acknowledgments}
This publication was made possible through the support of a Grant from the John Templeton Foundation.  The opinions expressed in this publication are those of the authors and do not necessarily reflect the views of the John Templeton Foundation. 

We especially wish to thank Mario Gattobigio for his many, informative conversations relevant to the Nuclear Physics aspects of our work.  It is also a pleasure to thank Mark Rostollan, of the American University of Paris, for his kind assistance in arranging sites for our collaborative research, when in Paris.
\end{acknowledgments}


\begin{thebibliography}{}
\bibitem{FootNote1}
{The present paper is the fifth in this series; the previous four papers, hereinafter denoted by 1, 2, 3, 4, are here listed with title, reference source, and a subject summary:
\begin{enumerate}
  \item "\emph{Gauge Invariant Summation of All QCD Gluon Exchanges}", Ref.~\cite{Fried2009_QCD1}, in which a new approach to analytic QCD is defined in the context of a quenched, eikonal approximation to $\mathrm{Q}\mathrm{Q}$ and $\mathrm{Q}\bar{\mathrm{Q}}$ scattering.
  \item "\emph{QCD and Effective Locality}", Ref.~\cite{Fried2010_QCD2}, in which the unexpected property of Effective Locality (EL) rigorously appears, demonstrating the validity of the gauge-invariant, non-perturbative results of [1]. A proof is given showing that these results hold without approximation and without exception.
  \item "\emph{Ideal vs. Realistic QCD}", Ref.~\cite{Fried2011_QCD3}, in which a rigorous application of Effective Locality shows that the neglect of basic imprecision of (bound) quark transverse coordinates, here denoted by "ideal" QCD, is untenable in a non-perturbative context, for it leads to an empty theory. A simple change to a phenomenological "realistic" QCD, redefined in its basic Lagrangian, removes this difficulty and sets the stage for future, sensible calculations.
  \item "\emph{Quark Binding Potentials}", Ref.~\cite{Fried2011_QCD4}, in which the above ideas are used to define quark-binding potentials for a model "pion" and "nucleon".  A double minimization technique provides an estimate of the pion mass, the ground state of a $\mathrm{Q}$-$\bar{\mathrm{Q}}$ system (in terms of a mass parameter $\mu$ of value fixed in the present paper); and one finds that the bound-state energy contained in the gluon fields is approximately three times as large as the sum of the $\mathrm{Q}$ and $\bar{\mathrm{Q}}$ rest-mass energies.
\end{enumerate}
}

%
%
%

%\bibitem[Reinhardt(1993)]{Reinhardt:1993}
%H. Reinhardt, K. Langfeld, and L. v. Smekal, \emph{Phys. Lett.} B\textbf{300}, (1993) 111.

%\bibitem{Ralf:2008}
%R. Hofmann, Int. J. Mod. Phys. {{A}}{\bf{20}}, (2006), 4123. Erratum-ibid. A {\bf{21}} (2006) 6515.

%\bibitem{DanielDF:2011}
%D.D. Ferrante, G.S. Guralnik, Z. Guralnik, C. Pehlevan. \emph{BOWN-HET-1611}

\bibitem{Fried2009_QCD1}
H. M. Fried, Y. Gabellini, T. Grandou and Y.-M. Sheu, Eur. Phys. J. \textbf{C65}, 395 (2010).

%
% arXiv
%
\bibitem{Fried2010_QCD2}
H. M. Fried, M. Gattobigio, T. Grandou and Y.-M. Sheu, arXiv:1003.2936 [hep-th].

\bibitem{Fried2011_QCD3}
H. M. Fried, T. Grandou and Y.-M. Sheu, arXiv:1103.4179 [hep-th].

\bibitem{Fried2011_QCD4}
H. M. Fried, Y. Gabellini, T. Grandou and Y.-M. Sheu, arXiv:1104.4663 [hep-th].

%\bibitem{Fried2011_QCD5}
%H. M. Fried, Y. Gabellini, T. Grandou and Y.-M. Sheu, in preparation.

\bibitem{bctg}
B. Candelpergher and T. Grandou, in preparation.

\bibitem{Fried2012_QCDII}
H. M. Fried, M. Gattobigio, T. Grandou and Y.-M. Sheu, in preparation.
%

\bibitem{Schwinger1951}
J. Schwinger, Phys. Rev. \textbf{82}, 664 (1951).

%\bibitem{Fradkin1966}
%E. S. Fradkin, Nucl. Phys. \textbf{76}, 588 (1966).

%\bibitem{Cheng-n-Wu1970}
%H. Cheng and T.T. Wu, Phys. Rev. Lett. \textbf{24}, (1970 ) 1456.

%\bibitem{bctg:2012}
%B. Candelpergher and T. Grandou, work in completion.

%\bibitem{Halpern1977a}
%M. B. Halpern, Phys. Rev. D\textbf{16}, (1977) 1798.

%\bibitem{Halpern1977b}
%M. B. Halpern, Phys. Rev. D\textbf{16}, (1977) 3515.

%\bibitem{Guay:2004}
%A. Guay, {\it{Geometrical apects of local gauge symmetry}}, (2004), http://philsci-archive.pitt.edu/id/eprint/2133.

%\bibitem{Huang:1977}
%K. Huang and D. R. Stump, Phys. Rev. Lett. \textbf{37}, (1976) 545; Phys. Rev. D \textbf{15}, (1977) 3660.

%\bibitem{hftg:2012}
%H. M. Fried, T. Grandou and Y.-M. Sheu, work in completion.

%\bibitem{11tg:2011}
%T. Grandou, {\it{On Some Technical Aspects of the QCD E?ective Locality Property}} ÒEleventh Workshop on Non-Perturbative QCDÓ, Juin 2011, B. M\"uller and C.I. Tan Editeurs, http://www.slac.stanford.edu/econf/C1106064/

%\bibitem{Ref-Machine}
%M.X. Etc. Phys.  etc..

\bibitem{Quantics JMLL}
F. Balibar, A. Laverne and J.M. Levy Leblond, {\it{ Quantique: El\'ements}},
\url{http://cel.archives-ouvertes.fr/docs/00/13/61/89/PDF/elem_5fev07.pdf}.

\bibitem{Fried2000}
H. M. Fried, Y. Gabellini, J. Avan, Eur. Phys. J. C\textbf{13}, 699 (2000).

%\bibitem{levy}
%For example, see \url{http://en.wikipedia.org/wiki/Stable_distribution}

\bibitem{Islam}
M. Islam, \textit{Proton Structure and prediction of pp elastic scattering at 7 TeV}, Proceedings of the 11th Workshop on Non-Perturbative QCD, Paris, June 2011.

\bibitem{Jastrow1951}
R. Jastrow, Phys. Rev. \textbf{81}, 664 (1951).  {A very nice fit to the shape of the potential of Fig.~\ref{Fig:3} of the present paper is the average of the singlet and triplet potentials of Fig. 1 of this reference.}


\bibitem{HMF1}
H. M. Fried, \textit{Functional Methods and Models in Quantum Field Theory} (The MIT Press, Cambridge, MA 1972)

\bibitem{HMF2}
H. M. Fried, \textit{Basics of Functional Methods and Eikonal Models} (Editions Fronti\`{e}res, Gif-sur-Yvette Cedex, France 1990)

\bibitem{HMF3}
H. M. Fried, \textit{Green's Functions and Ordered Exponentials} (Cambridge University Press, Cambridge 2002)

%\bibitem{Lapidus2000}
%G. W. Johnson and M. L. Lapidus, \textit{The Feynman Integral and Feynman's Operational Calculus} (Oxford University Press, Oxford 2000)

%\bibitem{GR}
%I. S. Gradshteyn and I. M. Ryzhik, \textit{Table of Integrals, Series and Products} (Academic Press, London 1994) formula 8.253.1

%\bibitem{Schiff}
%L. I. Schiff, \textit{Quantum Mechanics, 3rd Ed.} (McGraw-Hill, 1968)

\bibitem{Cheng-n-Wu1987}
H. Cheng and T. T. Wu, \textit{Expanding Protons: Scattering at High Energies} (MIT Press, Cambridge, MA, 1987)

%\bibitem{YMS2008}
%Y.-M. Sheu, "{\it{Finite-Temperature Quantum Electrodynamics: General Theory and Bloch-Nordsieck Estimates of Fermion Damping in a Hot Medium}}", PhD Thesis, Brown University, May 2008.

%\bibitem {Zee:2010}
%A. Zee, \textit{Quantum Field Theory in a Nutshell} (Princeton University Press, Princeton 2010)

%
% etc
%
\bibitem{Note}
$w_{0}= (0_{0},\vec{y}_{\perp}, 0_{\mathrm{L}})$, correcting the expression given in \cite{Fried2009_QCD1}

%\bibitem{LeonardCohen}
%Adapted from the poem "\emph{Anthem}" by Leonard Norman Cohen

\bibitem{FootNote3}
{More precisely, the exponential of the gauge-dependent gluon propagator standing to the left of the linkage operator retains its exact form in the course of any calculation devoted solely to gluons, but all radiative corrections to that propagator are gauge invariant, by construction, because the Fradkin representation for $\mathbf{L}[A]$ is invariant under the full group of SU(3) transformations. And "gluon bundles", comprising an infinite number of gluons exchanged between quarks and/or antiquarks are gauge invariant in the extreme sense that all relevant, gauge-dependent gluon propagators cancel out of their final Gaussian evaluation.}

\bibitem{Eikonal}
{The history and genesis of the eikonal model in High-Energfy Physics may be found in many papers and several books which explain and reference those papers, \emph{e.g.}, Ref.~\cite{Cheng-n-Wu1987} and \cite{HMF2}. A detailed derivation of the eikonal limit for non-Abelian theories may be found in Appendix B, of Ref.~\cite{Fried2000}}

\bibitem{ClusterExp}
{Detailed derivations may be found in \cite{HMF2} and \cite{HMF3}.}
%
%
\end{thebibliography}
\end{document}